\documentclass[epj]{svjour}

\usepackage[cp1250]{inputenc} 
\usepackage{a4wide}
\usepackage{authblk}
\usepackage{amsfonts}
\usepackage{amsmath}
\usepackage{amssymb}
\usepackage{booktabs}
\usepackage{epsfig}
\usepackage[colorlinks]{hyperref}
\hypersetup{citecolor=blue,linkcolor= blue}
\begin{document}
\title{Measuring capital market efficiency: Long-term memory, fractal dimension and approximate entropy}
\author{Ladislav Kristoufek\inst{1,2} \and Miloslav Vosvrda\inst{1,2}}
\institute{Institute of Information Theory and Automation, Academy of Sciences of the Czech Republic, Pod Vodarenskou Vezi 4, 182 08, Prague, Czech Republic, EU
 \and Institute of Economic Studies, Faculty of Social Sciences, Charles University in Prague, Opletalova 26, 110 00, Prague, Czech Republic, EU
}
\date{Received: date / Revised version: date}
%
\abstract{We utilize long-term memory, fractal dimension and approximate entropy as input variables for the Efficiency Index [Kristoufek \& Vosvrda (2013), Physica A 392]. This way, we are able to comment on stock market efficiency after controlling for different types of inefficiencies. Applying the methodology on 38 stock market indices across the world, we find that the most efficient markets are situated in the Eurozone (the Netherlands, France and Germany) and the least efficient ones in the Latin America (Venezuela and Chile).}

%
\authorrunning{Kristoufek \& Vosvrda}
\titlerunning{Measuring capital market efficiency}
\maketitle

\section{Introduction}

Efficient markets hypothesis (EMH) is one of the cornerstones of the modern financial economics. Since its introduction in 1960s \cite{Fama1965,Samuelson1965,Fama1970}, EMH has been a controversial topic. Nonetheless, the theory still remains a stable part of the classical financial economics. Regardless of its definition via a random walk \cite{Fama1965} or a martingale \cite{Samuelson1965}, the main idea of EMH is that risk-adjusted returns cannot be systematically predicted and there can be no long-term profits above the market profits assuming the same risk. The EMH definition is also tightly connected with a notion of rational homogenous agents and Gaussian distribution of returns. Both these assumptions have been widely disregarded in the literature \cite{Cont2001}.

In the econophysics literature, EMH has been most frequently studied with respect to the correlation structure of the series. There are several papers ranking various financial markets with respect to their efficiency. Research group around Di Matteo \cite{DiMatteo2003,DiMatteo2005,DiMatteo2007} finds that the correlations structure of various assets (stocks, exchange rates and interest rates) is connected to the development of the specific countries and stock markets. The importance of long-term memory and multifractality in the financial series is then further discussed in the subsequent research of the group \cite{Barunik2012,Morales2012,Morales2013}. In the series of papers, Cajueiro \& Tabak \cite{Cajueiro2004,Cajueiro2004a,Cajueiro2004b,Cajueiro2005} study the relationship between the long-term memory parameter $H$ and development stages of the countries' economy. Both groups find interesting results connecting persistent (long-term correlated) behavior to the least developed markets but also anti-persistent behavior for the most developed ones. Lim \cite{Lim2007} investigates how the ranking of stock markets based on Hurst exponent evolves in time and reports that the behavior can be quite erratic. Zunino \textit{et al.} \cite{Zunino2010} utilize entropy to rank stock markets to show that the emergent/developing markets are indeed less efficient than the developed ones. Even though the ranking is provided in these studies, the type of memory taken into consideration (either long-term memory or entropy/complexity) is limited and treated separately. 

In this paper, we utilize the Efficiency Index proposed by Kristoufek \& Vosvrda \cite{Kristoufek2013} incorporating long-term memory, fractal dimension and entropy to control for various types of correlations and complexity using a single measure. Basing the definition of the market efficiency simply on no correlation structure, we can state the expected values of long-term memory, fractal dimension and entropy for the efficient market to construct an efficiency measure based on a distance from the efficient market state. Introduction of the entropy measure into the Efficiency Index is novel compared to the original one \cite{Kristoufek2013} and it substitutes the short-term memory effect of the Index which turned out to be a rather weak component of the Index. Short-term memory inefficiency is still controlled for by inclusion of the fractal dimension. As it turns out, the inclusion of the entropy measure has a strong effect on the final efficiency ranking. The procedure is applied on 38 stock indices from different parts of the world and we show that the most efficient markets are indeed the most developed ones -- the Western European markets and the US markets -- and majority of the least efficient ones are situated in the Latin America and South-East Asia. 

The paper is structured as follows. In Section 2, we provide brief description of used methodology focusing on long-term memory, fractal dimension, entropy and efficiency measure. Section 3 introduces the dataset and describes the results. Section 4 concludes.

\section{Methodology}

\subsection{Long-term memory}

Long-term memory (long-range dependence) is usually characterized in time domain by a power-law decay of autocorrelation function and in frequency domain by a power-law divergence of spectrum close to the origin. More specifically, the autocorrelation function $\rho(k)$ with lag $k$ of a long-range correlated process decays as $\rho(k) \propto k^{2H-2}$ for $k\rightarrow +\infty$, and the spectrum $f(\lambda)$ with frequency $\lambda$ of a long-range correlated process diverges as $f(\lambda)\propto \lambda^{1-2H}$ for $\lambda \rightarrow 0+$. The characteristic parameter of the long-term memory Hurst exponent $H$ ranges between $0 \le H < 1$ for stationary processes. The breaking value of 0.5 indicates no long-term memory so that the autocorrelations decay rapidly (exponentially or faster). For $H>0.5$, the series is persistent with strong positive correlations characteristic by a trend-like behavior while still remaining stationary. For $H<0.5$, the series is anti-persistent and it switches the direction of increments more frequently than a random process does. 

There are many different estimators of the long-term memory parameter $H$ in both frequency and time domains \cite{Taqqu1995,Taqqu1996,Teverovsky1999,Barunik2010}. However, the estimators are usually affected by short-term memory bias \cite{Teverovsky1999,Kristoufek2012}, distributional properties \cite{Barunik2010,Kristoufek2010a,Kristoufek2012} or finite-size effect \cite{Weron2002,Couillard2005,Lennartz2009,Zhou2012} causing the estimates to have rather wide confidence intervals for these specific cases. To avoid these issues, we utilize two estimators from the frequency domain -- the local Whittle and GPH estimators -- which are appropriate for rather short financial series with a possible weak short-term memory \cite{Taqqu1995,Taqqu1996} and moreover, they have well-defined asymptotic properties -- consistency and asymptotic normality. Efficiency Index is then based on these estimators of Hurst exponent $H$.

\subsubsection*{Local Whittle estimator}
The local Whittle estimator \cite{Robinson1995a} is a semi-parametric maximum likelihood estimator -- the method utilizes a likelihood function of K\"{u}nsch \cite{Kunsch1987} and focuses only on a part of spectrum near the origin. The periodogram $I(\lambda_j)=\frac{1}{T}\sum_{t=1}^{T}{\exp(-2\pi i t\lambda_j)x_t}$ is utilized as an estimator of the spectrum of a series $\{x_t\}$ with $j=1,2,\ldots,m$ where $m\le T/2$ and $\lambda_j=2\pi j/T$. Assuming that series is indeed long-range correlated with $0\le H < 1$, the local Whittle estimator is defined as
\begin{equation}
\label{eq:LWX}
\widehat{H}=\arg \min_{0\le H <1} R(H),
\end{equation} 
where 
\begin{equation}
\label{eq:LWX_R}
R(H)=\log\left(\frac{1}{m}\sum_{j=1}^m{\lambda_j^{2H-1}I(\lambda_j)}\right)-\frac{2H-1}{m}\sum_{j=1}^m{\log \lambda_j}.
\end{equation}
The local Whittle estimator is consistent and asymptotically normal, specifically 
\begin{equation}
\sqrt{m}(\widehat{H}-H^0) \rightarrow_d N(0,1/4).
\end{equation}

\subsubsection*{GPH estimator}

The GPH estimator, named after Geweke \& Porter-Hudak \cite{Geweke1983}, is based on a full functional specification of the underlying process as the fractional Gaussian noise implying a specific spectral form:
\begin{equation}
\log f(\lambda) \propto -(H-0.5)\log [4\sin^2(\lambda/2)]
\end{equation}
Again, the spectrum needs to be estimated using the periodogram so that Hurst exponent is estimated using the least squares method to the following equation:
\begin{equation}
\label{GPH}
\log I(\lambda_j) \propto -(H-0.5)\log [4\sin^2(\lambda_j/2)]
\end{equation}

The GPH estimator is consistent and asymptotically normal \cite{Beran1994}, specifically 
\begin{equation}
\sqrt{T}(\widehat{H}-H^0) \rightarrow_d N(0,\pi^2/6).
\end{equation}
Asymptotically, the GPH estimator is thus infinitely more efficient than the local Whittle estimator. However, this holds only if the true underlying process is indeed the fractional Gaussian noise. In financial series, this is frequently not the case and the processes are mostly combinations of short-term and long-term memory processes. The GPH estimator then becomes biased. To overcome this issue, we base the GPH estimator only on a part of the spectrum (periodogram) close to the origin to avoid the short-term memory bias. The regression in Eq. \ref{GPH} is then not run on all $\lambda_j$ frequencies but only for a part based on the same parameter $m$ as for the local Whittle estimator.

\subsection{Fractal dimension}

Fractal dimension $D$ is a measure of roughness of the series and can be taken as a measure of local memory of the series \cite{Kristoufek2013}. For a univariate series, it holds that $1<D\le 2$. For self-similar processes, the fractal dimension is connected to the long-term memory of the series so that $D+H=2$. This can be attributed to a perfect reflection of a local behavior (fractal dimension) to a global behavior (long-term memory). However, the relation usually does not hold perfectly for the financial series so that both $D$ and $H$ give different insights into the dynamics of the series. In general, $D=1.5$ holds for a random series with no local trending or no local anti-correlations. For a low fractal dimension $D<1.5$, the series is locally less rough and thus resembles a local persistence. Reversely, a high fractal dimension $D>1.5$ is characteristic for rougher series with local anti-persistence. For purposes of the Efficiency Index, we utilize Hall-Wood and Genton estimators \cite{Gneiting2004,Gneiting2010}.

\subsubsection*{Hall-Wood estimator}

Hall-Wood estimator \cite{Hall1993} is based on box-counting procedure and utilizes scaling of absolute deviations between steps. Formally, let's have
\begin{equation}
\widehat{A(l/n)}=\frac{l}{n}\sum_{i=1}^{\lfloor n/l \rfloor}{|x_{il/n}-x_{(i-1)l/n}|}
\end{equation}
which represents these absolute deviations for the series of length $n$ within boxes of size $l$. Based on the definition of the fractal dimension \cite{Gneiting2004,Gneiting2010}, the Hall-Wood estimator is given by
\begin{equation}
\widehat{D_{HW}}=2-\frac{\sum_{l=1}^{L}{(s_l-\bar{s})\log(\widehat{A(l/n)})}}{\sum_{l=1}^{L}{(s_l-\bar{s})^2}}
\end{equation}
where $L \ge 2$, $s_l=\log(l/n)$ and $\bar{s}=\frac{1}{L}\sum_{l=1}^{L}{s_l}$. Using $L=2$ as suggested by Hall \& Wood \cite{Hall1993} to minimize bias, we get
\begin{equation}
\widehat{D_{HW}}=2-\frac{\log\widehat{A(2/n)}-\log\widehat{A(1/n)}}{\log2}.
\end{equation}

\subsubsection*{Genton estimator}

Genton estimator is a method of moments estimator \cite{Gneiting2004,Gneiting2010} based on the robust estimator of variogram of Genton \cite{Genton1998}. Defining the variogram as
\begin{equation}
\widehat{V_2(l/n)}=\frac{1}{2(n-l)}\sum_{i=l}^{n}{(x_{i/n}-x_{(i-l)l/n})^2},
\end{equation}
we get the Genton estimator as
\begin{equation}
\widehat{D_{G}}=2-\frac{\sum_{l=1}^{L}{(s_l-\bar{s})\log(\widehat{V_2(l/n)})}}{2\sum_{l=1}^{L}{(s_l-\bar{s})^2}}
\end{equation}
where again $L \ge 2$, $s_l=\log(l/n)$ and $\bar{s}=\frac{1}{L}\sum_{l=1}^{L}{s_l}$. Using $L=2$ \cite{Davies1999,Zhu2002} to decrease the bias again, we get
\begin{equation}
\widehat{D_{G}}=2-\frac{\log\widehat{V_2(2/n)}-\log\widehat{V_2(1/n)}}{2\log2}.
\end{equation}

\subsection{Approximate entropy}

Entropy can be taken as a measure of complexity of the system. The systems with high entropy can be characterized by no information and are thus random and reversely, the series with low entropy can be seen as deterministic \cite{Pincus2004}. The efficient market can be then seen as the one with maximum entropy and the lower the entropy, the less efficient the market is. For purposes of the Efficiency Index, we need an entropy measure which is bounded. Therefore, we utilize the approximate entropy introduced by Pincus \cite{Pincus1991}.

For each $i$ in $1\le i \le T-m+1$, we define
\begin{equation}
C_i^m(r)=\frac{\sum_{j=1}^{T-m+1}{\textbf{1}_{d[i,j]\le r}}}{T-m+1}
\end{equation}
where $\textbf{1}_{\bullet}$ is a binary indicator function equal to 1 if the condition in $\bullet$ is met and 0 otherwise and where
\begin{equation}
d[i,j]=\max_{k=1,2,\ldots,m}(|x_{i+k-1}-u_{j+k-1}|).
\end{equation}
$C_i^m(r)$ can be thus seen as a measure of auto-correlation as it is based on a maximum distance between lagged series. Averaging $C_i^m(r)$ across $i$ yields
\begin{equation}
C^m(r)=\frac{1}{T-m+1}\sum_{i=1}^{T-m+1}{C_i^m(r)}
\end{equation}
which is connected to the correlation dimension
\begin{equation}
\beta_m=\lim_{r\rightarrow 0}\lim_{T\rightarrow +\infty} \frac{\log C^m(r)}{\log r} 
\end{equation}
which is in turn treated as a measure of entropy and complexity of the series \cite{Pincus1991}. $\beta_m$ ranges between 0 (completely deterministic) and 1 (completely random).

\subsection{Capital market efficiency measure}

According to Kristoufek \& Vosvrda \cite{Kristoufek2013,Kristoufek2014}, the Efficiency Index (EI) is defined as

\begin{equation}
EI=\sqrt{{\sum_{i=1}^n{\left(\frac{\widehat{M_i}-M_i^{\ast}}{R_i}\right)^2}}},
\end{equation}
where $M_i$ is the $i$th measure of efficiency, $\widehat{M_i}$ is an estimate of the $i$th measure, $M_i^{\ast}$ is an expected value of the $i$th measure for the efficient market and $R_i$ is a range of the $i$th measure. In words, the efficiency measure is simply defined as a distance from the efficient market specification based on various measures of the market efficiency. In our case, we consider three measures of market efficiency -- Hurst exponent $H$ with an expected value of 0.5 for the efficient market ($M_H^{\ast}=0.5$), fractal dimension $D$ with an expected value of 1.5 ($M_D^{\ast}=1.5$) and the approximate entropy with an expected value of 1 ($M_{AE}^{\ast}=1$). The estimate of Hurst exponent is taken as an average of estimates based on GPH and the local Whittle estimators. The estimate of the fractal dimension is again taken as an average of the estimates based on the Hall-Wood and Genton methods. For the approximate entropy, we utilize the estimate described in the corresponding section. However, as the approximate entropy ranges between 0 (for completely deterministic market) and 1 (for random series), we need to rescale its effect, i.e. we use $R_{AE}=2$ for the approximate entropy and $R_H=R_D=1$ for the other two measures so that the maximum deviation from the efficient market value is the same for all measures.

\section{Application and discussion}

We analyze 38 stock indices from various locations -- the complete list is given in Tab. \ref{tab1} -- between January 2000 and August 2011. Various phases of the market behavior -- DotCom bubble, bursting of the bubble, stable growth of 2003-2007 and the current financial crisis -- are thus covered in the analyzed period. The indices cover stock markets in both North and Latin Americas, Western and Eastern Europe, Asia and Oceania so that markets at various levels of development are included in the study. Tab. \ref{tab2} summarizes the basic descriptive statistics of the analyzed indices -- the returns are asymptotically stationary (according to the KPSS test), leptokurtic and returns of majority of the indices are negatively skewed.

\begin{figure}[htbp]
\center
\begin{tabular}{cc}
\includegraphics[width=2.9in]{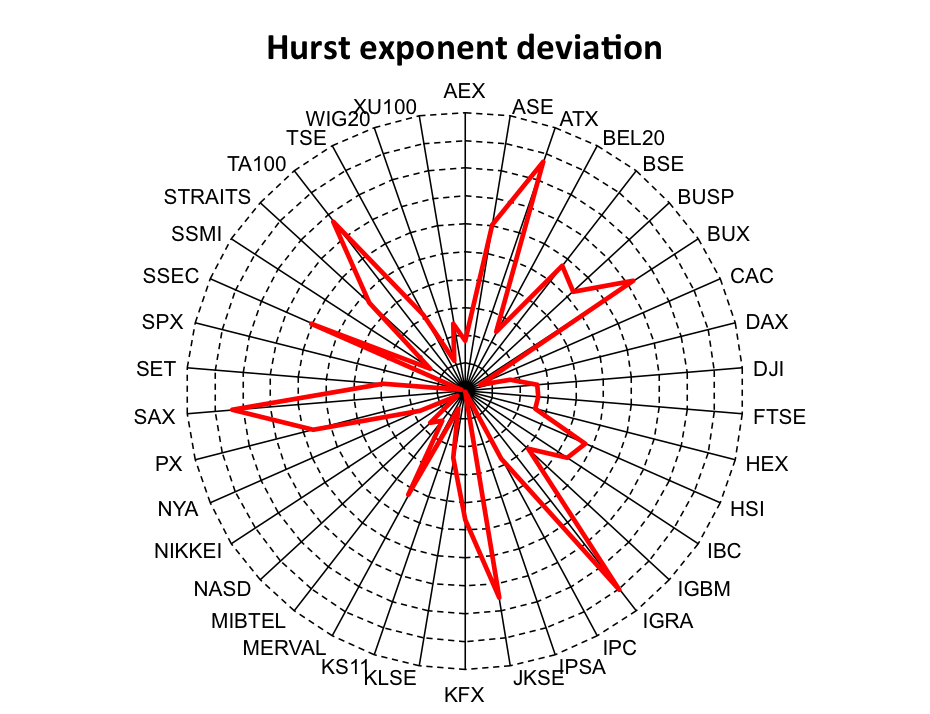}&\includegraphics[width=2.9in]{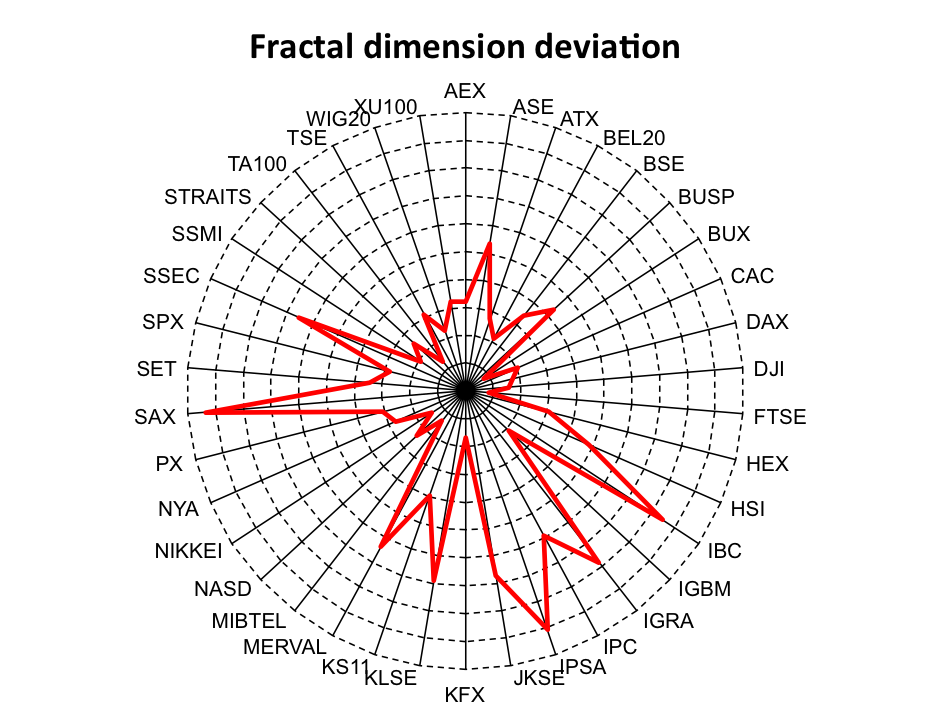}\\
\includegraphics[width=2.9in]{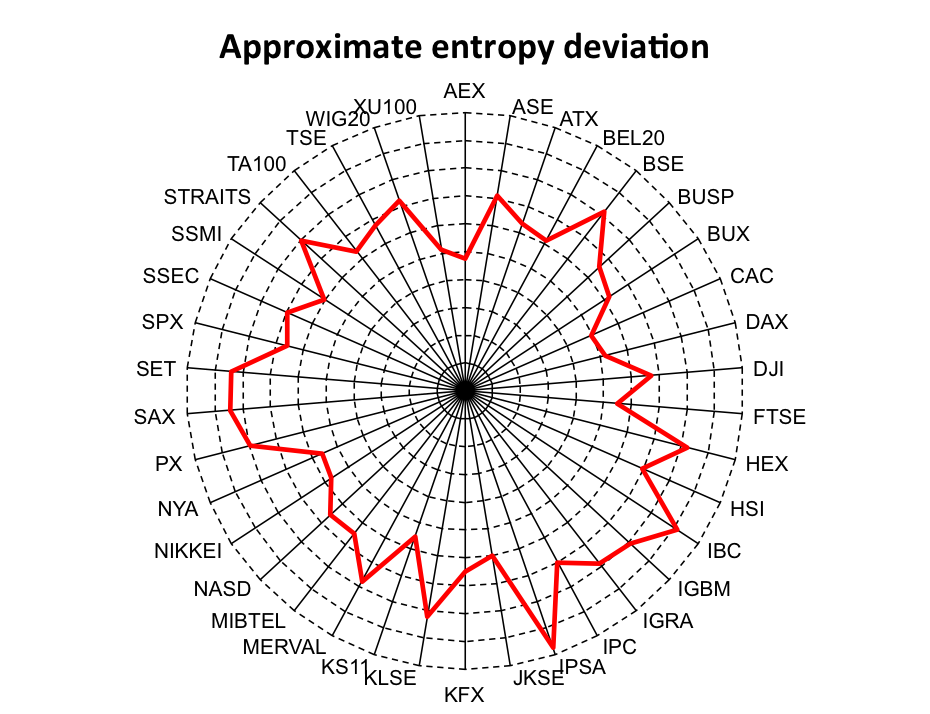}&\includegraphics[width=2.9in]{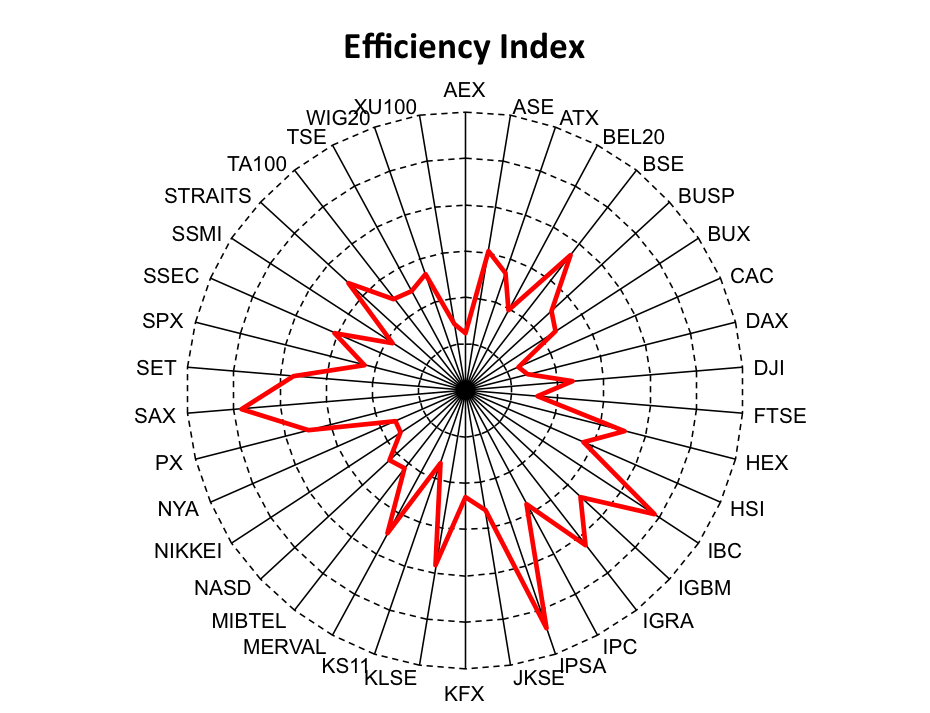}\\
\end{tabular}
\caption{\footnotesize\textit{Hurst exponent, fractal dimension, approximate entropy and efficiency index for analyzed indices.} The centers of the circle represent no deviation from the efficient market both for the specific deviations and for the Efficiency Index. The further the red line is from the center, the higher the deviation. The figures are rescaled to make the results more evident. From the Efficiency Index, we find that the Slovakian SAX, Venezuelan IBC, and Chilean IPSA are the least efficient markets whereas the Dutch AEX, French CAC and German DAX are the most efficient ones.   \label{Figures}}
\end{figure}

Let us now turn to the results. In Fig. \ref{Figures}, all the results are summarized graphically. For the utilized three measures -- Hurst exponent, fractal dimension and approximate entropy -- we present the absolute deviations from the expected values of the efficient market for comparison. For the Hurst exponent estimates, we observe huge diversity -- between practically zero (for IPSA of Chile) and 0.18 (for Peruvian IGRA). Interestingly, for some of the most developed markets, we observe Hurst exponents well below 0.5 (Tab. \ref{tab3} gives the specific estimates) which is, however, in hand with results of other authors \cite{DiMatteo2005,DiMatteo2007}. The results for the fractal dimension again vary strongly across the stock indices. The highest deviation is observed for the Slovakian SAX (0.19) and the lowest for the FTSE of the UK (0.02). In Tab. \ref{tab3}, we observe that apart from FTSE, all the other stock indices possess the fractal dimension below 1.5 which indicates that the indices are locally persistent, i.e. in some periods, the indices experience significant positively autocorrelated behavior which is well in hand with expectations about the herding behavior during critical events. The approximate entropy estimates are more stable across indices compared to the previous two cases. The highest deviation from the expected value for the efficient market is observed for the Chilean IPSA (0.98) and the lowest for the Dutch AEX (0.48). Evidently, all the analyzed stock indices are highly complex as the approximate entropy is far from the ideal (efficient market) value of 1 and such complexity is not sufficiently covered by the other two applied measures. The inclusion of the approximate entropy into the Efficiency Index thus proves its worth. 

Putting the estimates of the three measures together, we get the Efficiency Index which is also graphically presented in Fig. \ref{Figures}. For the ranking of indices according to their efficiency, we present Tab. \ref{tab3}. The most efficient stock market turns out to be the Dutch AEX closely followed by the French CAC and the German DAX. We can observe that the most efficient markets are usually the EU (or rather Eurozone) countries followed by the US markets and other developed markets from the rest of the world -- Japanese NIKKEI, Korean KS11, Swiss SSMI. The least efficient part of the ranking is dominated by the Asian and the Latin American countries. At the very end, we have the Slovakian SAX, Venezuelan IBC and Chilean IPSA. The efficiency of the stock markets is thus strongly geographically determined which is connected to the stage of development of the specific markets.

To see the contribution of the separate parts of the Index to the overall ranking, we present Tab. \ref{tab4} where the rankings according to the Efficiency Index and its components are compared. Evidently, the overall ranking is tightly connected to the ranking according to the entropy measure. However, the correspondence is not perfect -- Spearman's rank correlation between the two is equal to 0.94. For the fractal dimension and long-term memory components, the rank correlations are 0.65 and 0.49, respectively. It thus turns out that the stock indices are highly complex and this complexity plays the main role in their potential inefficiency. It also makes good sense that the effect of entropy dominates the ones of the fractal dimension and the long-term memory. In practice, it is hard to believe that stock indices would be persistent as such persistence would be quickly arbitraged out by profit-seeking traders. The fact that the fractal dimension has a stronger effect on the overall inefficiency compared to the long-term memory component is well in hand with the properties of the fractal dimension which tracks local, short-lived, correlations which are present in the stock indices. However, such dominance of the entropy measure in the overall Efficiency Index does not discredit utility of the Index itself as it turns out that such dominance might be stock index specific -- the Efficiency Index including entropy applied on various commodity futures does not show such a strong position of entropy compared to the other measures \cite{Kristoufek2014}.

Compared to the other studies mentioned in the Introduction section, our study provides a broader picture of treating the capital market efficiency. Most importantly, majority of the efficiency ranking studies focus on the long-term memory characteristics of the capital markets \cite{DiMatteo2003,DiMatteo2005,DiMatteo2007,Cajueiro2004,Cajueiro2004a,Cajueiro2004b,Cajueiro2005}. However, we show that the persistence or anti-persistence of the series plays only a marginal role in the overall efficiency ranking. This is well in hand with the assumption that any significant autocorrelations are quickly arbitraged away by algorithmic trading and noise traders. Such short-term profit opportunities represented by short-lived significant autocorrelations are captured by the fractal dimension which is found to be a more important component of the Efficiency Index. The most important role is attributed to the entropy, which makes our results partly comparable with the ones of Zunino \textit{et al.} \cite{Zunino2010} where the French CAC, German DAX and Italian MIB30 are, respectively, detected as the most efficient ones compared to our most efficient triad of the Dutch AEX, French CAC and German DAX in a descending order. However, the dataset of the former study does not include the Dutch stock index. And even though the most efficient triplets are very alike, the rest of the ranking differs more which we attribute to more sources of inefficiencies taken into consideration by the Efficiency Index presented in this study.

\section{Conclusions}

We have utilized long-term memory, fractal dimension and approximate entropy as input variables for the Efficiency Index \cite{Kristoufek2012b,Kristoufek2013}. This way, we are able to comment on stock market efficiency after controlling for different types of inefficiencies. Applying the methodology on 38 stock market indices across the world, we find that the most efficient markets are situated in the Eurozone (the Netherlands, France and Germany) and the least efficient ones in the Latin America (Venezuela and Chile). The Efficiency Index thus well corresponds to the expectation that the stock market efficiency is connected to the development of the market.

\section*{Acknowledgements}

The research leading to these results has received funding from the European Union's Seventh Framework Programme (FP7/2007-2013) under grant agreement No. FP7-SSH-612955 (FinMaP) and the Czech Science Foundation project No. P402/12/G097 ``DYME -- Dynamic Models in Economics''.


\bibliographystyle{plain}
\bibliography{EMH}

\onecolumn

\begin{table}[htbp]
\centering
\caption{List of the analyzed indices}
\label{tab1}
\footnotesize
\begin{tabular}{||c|c|c||}
\toprule \toprule
Ticker&Index&Country\\
\midrule \midrule
AEX&Amsterdam Exchange Index&Netherlands\\
ASE&Athens Stock Exchange General Index&Greece\\
ATX&Austrian Traded Index&Austria\\
BEL20&Euronext Brussels Index&Belgium\\
BSE&Bombay Stock Exchange Index&India\\
BUSP&Bovespa Brasil Sao Paulo Stock Exchange Index&Brasil\\
BUX&Budapest Stock Exchange Index&Hungary\\
CAC&Euronext Paris Bourse Index&France\\
DAX&Deutscher Aktien Index&Germany\\
DJI&Dow Jones Industrial Average Index&USA\\
FTSE&Financial Times Stock Exchange 100 Index&UK\\
HEX&OMX Helsinki Index&Finland\\
HSI&Hang Seng Index&Hong-Kong\\
IBC&Caracas Stock Exchange Index&Venezuela\\
IGBM&Madrid Stock Exchange General Index&Spain\\
IGRA&Peru Stock Market Index&Peru\\
IPC&Indice de Precios y Cotizaciones&Mexico\\
IPSA&Santiago Stock Exchange Index&Chile\\
JKSE&Jakarta Composite Index&Indonesia\\
KFX&Copenhagen Stock Exchange Index&Denmark\\
KLSE&Bursa Malaysia Index&Malaysia\\
KS11&KOSPI Composite Index&South Korea\\
MERVAL&Mercado de Valores Index&Argentina\\
MIBTEL&Borsa Italiana Index&Italy\\
NASD&NASDAQ Composite Index&USA\\
NIKKEI&NIKKEI 225 Index&Japan\\
NYA&NYSE Composite Index&USA\\
PX&Prague Stock Exchange Index&Czech Republic\\
SAX&Slovakia Stock Exchange Index&Slovakia\\
SET&Stock Exchange of Thailand Index&Thailand\\
SPX&Standard \& Poor's 500 Index&USA\\
SSEC&Shanghai Composite Index&China\\
SSMI&Swiss Market Index&Switzerland\\
STRAITS&Straits Times Index&Singapore\\
TA100&Tel Aviv 100 Index&Israel\\
TSE&Toronto Stock Exchange TSE 300 Index&Canada\\
WIG20&Warsaw Stock Exchange WIG 20 Index&Poland\\
XU100&Instanbul Stock Exchange National 100 Index&Turkey\\
\bottomrule
\end{tabular}
\end{table}

\begin{table}[htbp]
\centering
\caption{Descriptive statistics for the analyzed indices}
\label{tab2}
\footnotesize
\begin{tabular}{||c|cccc|cc|cc||}
\toprule \toprule
Index&mean&min&max&SD&skewness&ex. kurtosis&KPSS&p-value\\
\midrule \midrule
AEX&-0.0003&-0.0959&0.1003&0.0157&-0.0183&6.1531&0.1084&$>0.05$\\
ASE&-0.0006&-0.1021&0.1343&0.0169&-0.0697&5.0812&0.3531&$>0.05$\\
ATX&0.0002&-0.1025&0.1202&0.0150&-0.3410&8.2241&0.3141&$>0.05$\\
BEL20&-0.0001&-0.0832&0.0933&0.0135&0.0694&6.7098&0.1381&$>0.05$\\
BSE&0.0004&-0.1181&0.1599&0.0170&-0.1630&6.2487&0.1900&$>0.05$\\
BUSP&0.0004&-0.1210&0.1368&0.0193&-0.0641&4.5410&0.1229&$>0.05$\\
BUX&0.0004&-0.1265&0.1318&0.0169&-0.1105&6.3117&0.2860&$>0.05$\\
CAC&-0.0002&-0.0947&0.1060&0.0154&0.0594&5.3189&0.0944&$>0.05$\\
DAX&-0.0001&-0.0887&0.1080&0.0159&0.0025&4.7729&0.1681&$>0.05$\\
DJI&0.0000&-0.0820&0.1051&0.0126&-0.0089&7.8817&0.0647&$>0.05$\\
FTSE&-0.0001&-0.0927&0.0938&0.0129&-0.1309&6.4856&0.1222&$>0.05$\\
HEX&-0.0003&-0.1441&0.1344&0.0193&-0.1933&5.2159&0.1886&$>0.05$\\
HIS&0.0001&-0.1770&0.1341&0.0166&-0.2283&12.5630&0.1306&$>0.05$\\
IBC&0.0008&-0.2066&0.1453&0.0155&-0.4151&25.8530&0.2665&$>0.05$\\
IGBM&-0.0001&-0.1875&0.1840&0.0153&0.0833&20.5300&0.1272&$>0.05$\\
IGRA&0.0008&-0.1144&0.1282&0.0147&-0.3550&10.3010&0.3896&$>0.05$\\
IPC&0.0005&-0.0727&0.1044&0.0144&0.0515&4.3402&0.1295&$>0.05$\\
IPSA&0.0007&-0.0717&0.1180&0.0108&-0.0140&10.7400&0.1663&$>0.05$\\
JKSE&0.0006&-0.1095&0.0762&0.0150&-0.6570&6.1905&0.3397&$>0.05$\\
KFX&0.0002&-0.1172&0.0950&0.0137&-0.2594&5.7183&0.0939&$>0.05$\\
KLSE&0.0002&-0.1122&0.0537&0.0092&-1.1810&15.4970&0.1591&$>0.05$\\
KS11&0.0002&-0.1212&0.1128&0.0174&-0.4309&4.5849&0.1617&$>0.05$\\
MERVAL&0.0006&-0.1295&0.1612&0.0214&-0.1235&5.6617&0.1006&$>0.05$\\
MIBTEL&0.0002&-0.0771&0.0683&0.0108&-0.3979&5.7820&0.4301&$>0.05$\\
NASD&-0.0002&-0.1029&0.1116&0.0175&-0.1624&3.9587&0.2958&$>0.05$\\
NIKKEI&-0.0003&-0.1211&0.1324&0.0158&-0.3633&7.3242&0.1252&$>0.05$\\
NYA&0.0002&-0.1023&0.1153&0.0140&-0.4233&10.5210&0.1514&$>0.05$\\
PX50&0.0003&-0.1619&0.1236&0.0154&-0.6011&15.4230&0.4121&$>0.05$\\
SAX&0.0007&-0.0882&0.0711&0.0120&-0.0481&6.5294&0.5215&$>0.05$\\
SET&0.0000&-0.2211&0.1058&0.0158&-1.8111&26.2170&0.2975&$>0.05$\\
SPX&-0.0001&-0.0947&0.1096&0.0134&-0.1842&8.1808&0.0958&$>0.05$\\
SSEC&0.0002&-0.1200&0.0903&0.0168&-0.2784&4.7064&0.1461&$>0.05$\\
SSMI&-0.0001&-0.0811&0.1079&0.0127&0.0331&6.2488&0.0918&$>0.05$\\
STRAITS&0.0000&-0.2685&0.1406&0.0137&-2.2597&56.9590&0.1989&$>0.05$\\
TA100&0.0003&-0.0734&0.0978&0.0141&-0.1535&3.2977&0.1157&$>0.05$\\
TSE&0.0001&-0.0979&0.0937&0.0122&-0.6630&8.9915&0.0782&$>0.05$\\
WIG20&0.0004&-0.0886&0.3322&0.0185&2.6452&52.0680&0.1909&$>0.05$\\
XU100&0.0004&-0.1334&0.1749&0.0230&0.0039&4.5896&0.1105&$>0.05$\\
\bottomrule
\end{tabular}
\end{table}

\begin{table}[htbp]
\centering
\caption{Ranked stock indices according to the Efficiency Index}
\label{tab3}
\footnotesize
\begin{tabular}{||cc|ccc|c||}
\toprule \toprule
Index&Country&Hurst exponent&Fractal dimension&Approximate entropy&Efficiency index\\
\midrule \midrule
AEX&Netherlands&0.5358&1.4356&0.5246&0.0619\\
CAC&France&0.5118&1.4592&0.5059&0.0628\\
DAX&Germany&0.5334&1.4646&0.4807&0.0698\\
XU100&Turkey&0.5493&1.4350&0.4870&0.0724\\
FTSE&UK&0.4470&1.5171&0.4500&0.0787\\
NYA&USA&0.5348&1.4457&0.4418&0.0821\\
NIKKEI&Japan&0.5063&1.4716&0.4285&0.0825\\
KS11&South Korea&0.5137&1.4204&0.4473&0.0829\\
SSMI&Switzerland&0.5297&1.4617&0.3983&0.0929\\
BEL20&Belgium&0.5481&1.4574&0.3869&0.0981\\
MIBTEL&Italy&0.5267&1.4728&0.3525&0.1063\\
NASD&USA&0.5340&1.4526&0.3428&0.1114\\
SPX&USA&0.5026&1.4437&0.3405&0.1119\\
KFX&Denmark&0.5927&1.4665&0.3516&0.1148\\
DJI&USA&0.4477&1.4685&0.3284&0.1165\\
BUX&Hungary&0.6448&1.4844&0.3811&0.1170\\
TSE&Canada&0.5626&1.4375&0.3272&0.1210\\
TA100&Israel&0.6536&1.4739&0.3648&0.1251\\
BUSP&Brazil&0.6055&1.4142&0.3435&0.1262\\
JKSE&Indonesia&0.6505&1.3657&0.3986&0.1311\\
WIG20&Poland&0.5232&1.4545&0.2790&0.1326\\
ATX&Austria&0.6744&1.4455&0.3669&0.1336\\
HSI&Hong-Kong&0.5945&1.4033&0.3033&0.1396\\
IPC&Mexico&0.5550&1.3817&0.2991&0.1398\\
ASE&Greece&0.6210&1.3926&0.2911&0.1518\\
SSEC&China&0.6205&1.3698&0.3019&0.1533\\
IGBM&Spain&0.5615&1.4581&0.1912&0.1691\\
STRAITS&Singapore&0.5937&1.4500&0.2027&0.1702\\
PX&Czech Rep&0.6124&1.4386&0.2053&0.1743\\
MERVAL&Argentina&0.5850&1.3729&0.2225&0.1745\\
HEX&Finland&0.5524&1.4385&0.1747&0.1768\\
BSE&India&0.6139&1.4313&0.1842&0.1841\\
SET&Thailand&0.5591&1.4311&0.1590&0.1851\\
KLSE&Malaysia&0.5489&1.3620&0.1773&0.1906\\
IGRA&Peru&0.6806&1.3435&0.2160&0.2108\\
SAX&Slovakia&0.6673&1.3132&0.1534&0.2421\\
IBC&Venezuela&0.5881&1.3308&0.0890&0.2439\\
IPSA&Chile&0.4997&1.3187&0.0239&0.2711\\
\bottomrule
\end{tabular}
\end{table}

\begin{table}[htbp]
\centering
\caption{Ranking of the indices according to the components}
\label{tab4}
\footnotesize
\begin{tabular}{||cc|c|ccc||}
\toprule \toprule
Index&Country&Efficiency Index&Hurst exponent&Fractal dimension&Approximate entropy\\
\midrule \midrule
AEX&Netherlands&1&12&22&1\\
CAC&France&2&4&10&2\\
DAX&Germany&3&9&8&4\\
XU100&Turkey&4&15&23&3\\
FTSE&UK&5&18&2&5\\
NYA&USA&6&11&16&7\\
NIKKEI&Japan&7&3&5&8\\
KS11&South Korea&8&5&26&6\\
SSMI&Switzerland&9&8&9&10\\
BEL20&Belgium&10&13&12&11\\
MIBTEL&Italy&11&7&4&15\\
NASD&USA&12&10&14&18\\
SPX&USA&13&2&18&19\\
KFX&Denmark&14&25&7&16\\
DJI&USA&15&16&6&20\\
BUX&Hungary&16&33&1&12\\
TSE&Canada&17&22&21&21\\
TA100&Israel&18&35&3&14\\
BUSP&Brazil&19&28&27&17\\
JKSE&Indonesia&20&34&33&9\\
WIG20&Poland&21&6&13&26\\
ATX&Austria&22&37&17&13\\
HSI&Hong-Kong&23&27&28&22\\
IPC&Mexico&24&19&30&24\\
ASE&Greece&25&32&29&25\\
SSEC&China&26&31&32&23\\
IGBM&Spain&27&21&11&31\\
STRAITS&Singapore&28&26&15&30\\
PX&Czech Rep&29&29&19&29\\
MERVAL&Argentina&30&23&31&27\\
HEX&Finland&31&17&20&34\\
BSE&India&32&30&24&32\\
SET&Thailand&33&20&25&35\\
KLSE&Malaysia&34&14&34&33\\
IGRA&Peru&35&38&35&28\\
SAX&Slovakia&36&36&38&36\\
IBC&Venezuela&37&24&36&37\\
IPSA&Chile&38&1&37&38\\
\bottomrule
\end{tabular}
\end{table}

\end{document}